# Iterative annotation to ease neural network training: Specialized machine learning in medical image analysis

Brendon Lutnick[1], Brandon Ginley[1], Darshana Govind[1], Sean D. McGarry[2], Peter S. LaViolette[2], Rabi Yacoub[3], Sanjay Jain[4], John E. Tomaszewski[1], Kuang-Yu Jen[5], and Pinaki Sarder[1*]

[1]Department of Pathology & Anatomical Sciences, SUNY Buffalo, USA. [2]Department of Radiology, Medical College of Wisconsin. [3]Department of Medicine, Nephrology, SUNY Buffalo. [4]Department of Medicine, Nephrology, Washington University School of Medicine. [5]Department of Pathology, University of California, Davis Medical Center.

[*]Address all correspondence to: Pinaki Sarder
Tel: (716)-829-2265, e-mail: pinakisa@buffalo.edu

**Abstract**

Neural networks promise to bring robust, quantitative analysis to medical fields, but adoption is limited by the technicalities of training these networks. To address this translation gap between medical researchers and neural networks in the field of pathology, we have created an intuitive interface which utilizes the commonly used whole slide image (WSI) viewer, Aperio ImageScope (Leica Biosystems Imaging, Inc.), for the annotation and display of neural network predictions on WSIs. Leveraging this, we propose the use of a human-in-the-loop strategy to reduce the burden of WSI annotation. We track network performance improvements as a function of iteration and quantify the use of this pipeline for the segmentation of renal histologic findings on WSIs. More specifically, we present network performance when applied to segmentation of renal micro compartments, and demonstrate multi-class segmentation in human and mouse renal tissue slides. Finally, to show the adaptability of this technique to other medical imaging fields, we demonstrate its ability to iteratively segment human prostate glands from radiology imaging data.

**Introduction**

In the current era of artificial intelligence, robust automated image analysis is attained using supervised machine learning algorithms. This approach is gaining considerable ground in virtually every domain of data analysis, mainly under the advent of neural networks [2-5]. Neural networks are a broad range of algorithms which can take many different forms, but all are considered graphical models, whose nodes can be variably activated by a non-linear operation on the sum of their inputs [4, 6]. The connections between nodes are modulated by weights, which can be adjusted to dampen or amplify the power of contribution of that node to the output of the network. These weights can be iteratively tuned via back propagation so that the input of a particular type of data leads to a desired output (usually a classification of the data) [7]. Particularly useful for image analysis are convolutional neural networks (CNNs) [3, 4], a specialized subset of neural networks which leverage convolutional filters to learn spatial representations of image regions specific to the desired image classification. This allows high dimensional filtering operations to be learned automatically, a task which has traditionally been done through hand-engineering. Neural networks are problematic in certain applications, as they require significant amounts of annotated data (on the order of tens of thousands) in order to provide generalized high performance, yet their potential exceeds other machine learning techniques [8].

Work to ease the burden of data annotation is arguably as important as the generation of state-of-the-art network architectures, which without sufficient data are unusable [9, 10]. Many large-scale modern machine learning applications are indeed based on cleverly designed crowd sourced active learning pipelines, which in the era of constant firmware updates, comes in the form of human-in-the-loop training [11-13]. Initiated by low classification probabilities, machine learning applications such as automated teller machine character recognition, self-driving cars, and Facebook's automatic tagging, all rely on user refined training sets for fine tuning neural network applications post deployment [4]. These 'active learning' techniques require users to 'correct' the predictions of a network, therefore identifying gaps in network performance [14].

The adoption of neural networks to biological datasets has largely lagged behind adoption in computer science [15, 16]. While computational strategies for image analysis are seeing ever increasing translation to biological research, the late adoption of CNN-based methods for classification in biology is largely due to the lack of centrally curated and annotated training sets [17]. Due to the specialized nature of medical datasets, annotation by experts necessary for generation of training sets is less feasible than traditional datasets [18]. This issue creates challenges when trying to apply CNNs to medical imaging databases where domain-expert knowledge is required to perform image annotation, but domain-expert annotation is difficult to acquire because it is expensive, time consuming, labor intensive, and there are no technical mediums which enable easy transference of this information from clinical practice to training sets [19].

Despite of the above mentioned challenges in digital pathology, segmentation and classification of tissue slides by neural networks will not only aid clinical diagnosis based on current guidelines and practice, but will likely facilitate the creation of refined and improved future diagnostic guidelines using quantitative computational metrics. Additionally, neural networks can generate searchable data repositories [20], providing practicing clinicians and students access to collections of domain knowledge, such as labeled images and associated clinical outcomes that were not previously available [21-23]. While this end goal necessitates a combination of curated pathological datasets, machine learning classifiers [4], automatic anomaly detection [24, 25], and efficient searchable data hierarchies [22]; pipelines for creating easily viewable annotations on pathology images are a necessary first step. Towards this aim, we have developed an iterative interface between the successful semantic segmentation network DeepLab v2 [26] and the widely used WSI viewing software Aperio ImageScope [27], which we have termed Human A.I. Loop (H-AI-L) (Figure 1). Put simply, the algorithm converts annotated regions stored in XML format (provided in ImageScope) into image region masks. These masks are used to train the semantic segmentation network, whose predictions are converted back to XML format for display in ImageScope. This graphical display of network output is an

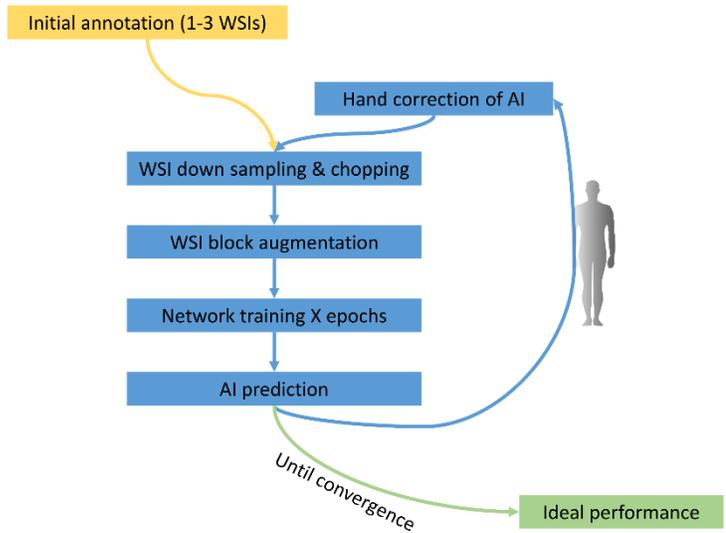

*Figure 1  Iterative Human A.I. Loop (H-AI-L) pipeline overview.*

Schematic representation of H-AI-L pipeline for training semantic segmentation of WSI. Several rounds of training are performed using human expert feedback in order to optimize ideal performance, resulting in improved efficiency in network training with limited numbers of initial annotated WSIs.

ideal visualization tool for segmentation predictions on WSI, with the ability to view the entire tissue slide, pan and zoom functionalities, as well as the efficient JPG2000 decompression [28] of WSI files provided by ImageScope. Using this open sourced pipeline, a supervising domain expert can correct the network predictions and initiate further training using the newly annotated data. This enables networks to be trained "on demand", or as the data is available.  Using H-AI-L, we are able to significantly reduce the annotation effort required to learn robust segmentations of large microscopy images [28]. Adaptation of this technique to other modes of medical imaging is highly feasible, which we demonstrate using MRI imaging data.

**Results**

To evaluate the utility of H-AI-L, we first quantified its performance and efficiency with histologic sections of kidney tissue, the first being glomerular localization in mouse kidney WSIs [5, 29-32]. This glomeruli segmentation network was trained for 5 iterations, using a combination of periodic acid-Schiff (PAS) and hematoxylin and eosin (H&E)-stained murine renal sections. For more data variation, streptozotocin (STZ) induced diabetic nephropathy [1, 33-35] murine data was included in iteration 4 (Table 1). To validate the performance of our network, we use 4 holdout WSIs, including one STZ induced WSI.

During the training process, we observed approximately 4 to 10-fold increases in average glomerular annotation speed between the initial and end iterations (Figure 2a). This represents time savings of 81.4%, 82%, and 72.7% for three annotators, annotator-1, -2, and -3, respectively, when compared to each annotator's baseline speed. This results in the prediction performance increase shown in Figure 2b, where the network reaches nearly perfect performance on a holdout dataset by annotation iteration 4. One side effect of using iterative annotation is the intuitive qualification of network performance it provides after each interaction; that is, an expert interacts with the network predictions after each training round, visualizing network biases and shortcomings on holdout data. Two examples of evolving network predictions are highlighted in Supplemental Figure 1.

In order to improve network prediction efficiency we designed a multi-resolution approach, which uses two segmentation networks: identifying hot spot regions at 1/16[th] scale before segmenting them at the highest resolution. This approach, which we call DeepZoom, obtains better F-measure (F1 score) [36, 37] (Figure 2b) versus a full resolution pass, as well as approximately 4.5-times faster predictions (Figure 2c). An overview of this method can be found in Supplemental Figure 2.

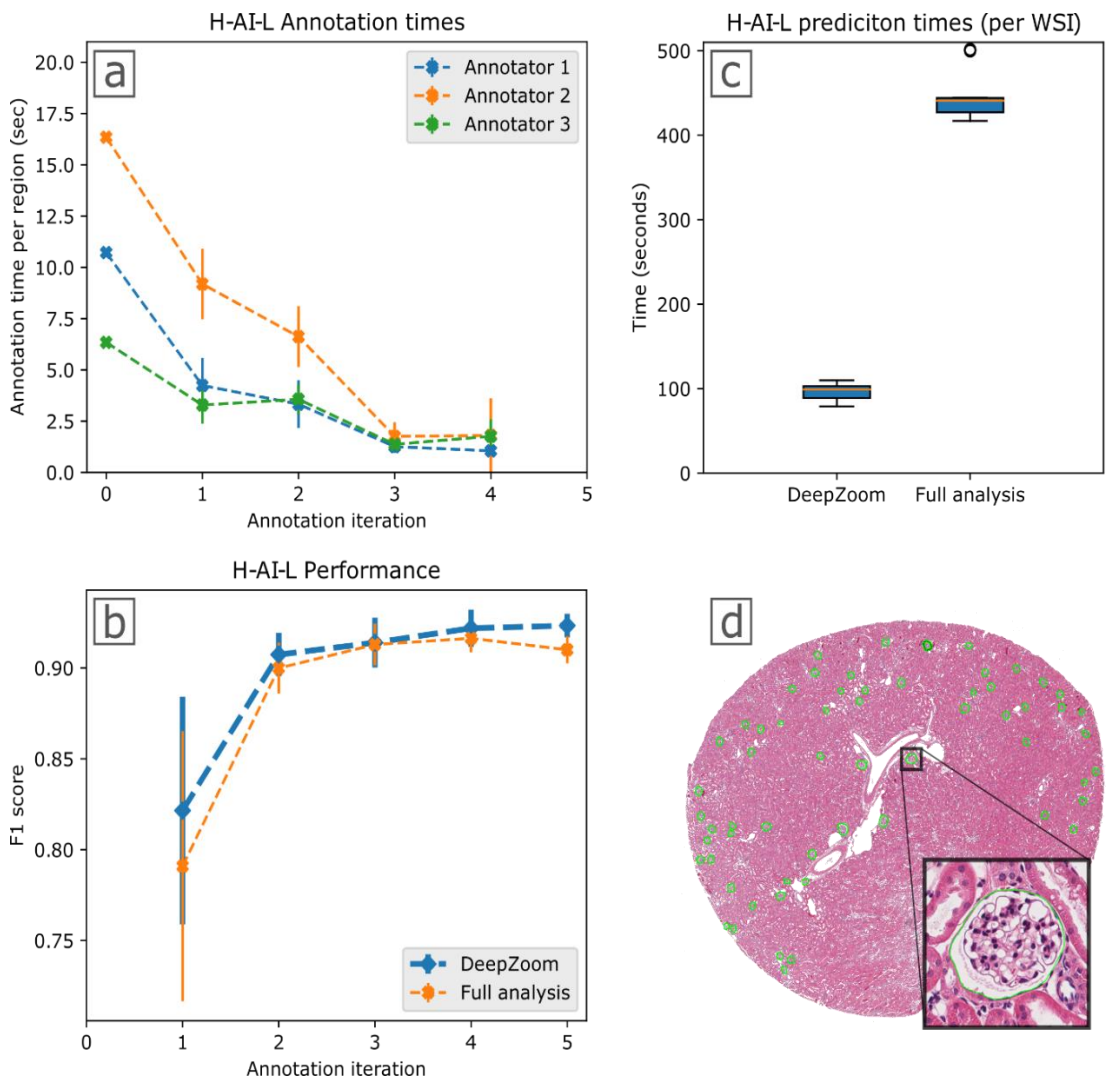

*Figure 2   H-AI-L pipeline performance: glomerular segmentation on holdout mouse WSI.*

**(a)** Annotation times per glomerulus as a function of annotation iteration. The 0[th] iteration was performed without preexisting predicted annotations, whereas subsequent iterations use network predictions as an initial annotation prediction that can be corrected by the annotator. **(b)** F1 score of glomerular segmentation of 4 holdout mouse renal WSIs as a function of training iteration. **(c)** Runtimes for glomerular segmentation prediction on holdout mouse renal WSIs using H-AI-L with DeepZoom (multi-resolution segmentation) versus full resolution segmentation. **(d)** Example of a mouse WSI with segmented glomeruli. Network predictions are outlined in green. Error bars indicate ±1 standard deviation.

| H-AI-L Data Set | | | | | | | |
|---|---|---|---|---|---|---|---|
| Annotation iteration | | 0 | 1 | 2 | 3 | 4 | Test |
| WSI added | | 1 | 2 | 4 | 6 | 4 | 4 |
| Total glomeruli | Normal | 32 | 84 | 86 | 418 | 0 | 138 |
| | STZ | 0 | 0 | 0 | 0 | 293 | 96 |

*Table 1   H-AI-L segmentation mouse WSI training and testing datasets.*

Mouse WSI training set used to train the glomerular segmentation network. Data presenting structural damage from streptozotocin (STZ) induced diabetes [1] was introduced in iteration 4. The test dataset included 3 normal and 1 STZ WSI.

Quantification of the performance achieved by our method in WSI is a challenge due to the imbalance between class distributions [38]. Therefore, we choose to report F-measure which considers both precision and recall (sensitivity) simultaneously [36], as specificity and accuracy are always high due to the large percentage of negative region with respect to the positive class. This is particularly important considering the performance characteristics of DeepZoom. During testing we found that DeepZoom trades segmentation sensitivity for increased precision, while outperforming full analysis overall with improved F1 score (Figure 2). This performance gap is due to a lower false positive rate achieved by DeepZoom as a result of the low resolution network pre-pass, which limits the amount of background region seen by the high resolution network. Overall, on four holdout WSIs, our network achieved its best performance after the $5^{th}$ iteration of training using DeepZoom with sensitivity 0.92 ± 0.02, specificity 0.99 ± 0.001, precision 0.93 ± 0.14, and accuracy 0.99 ± 0.001.

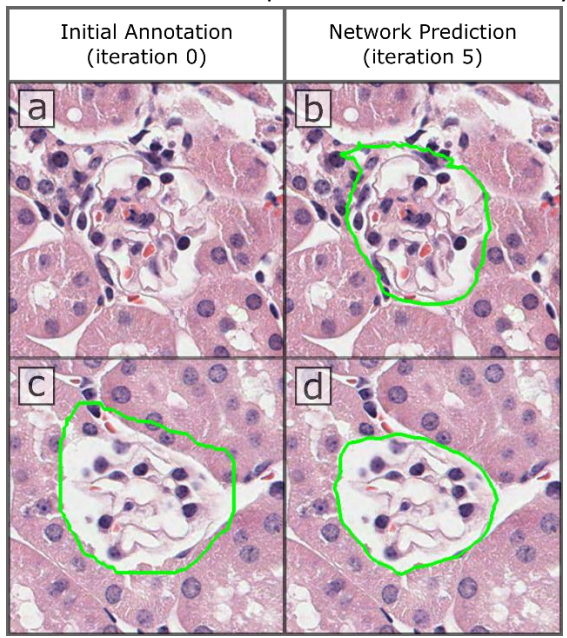

*Figure 3   H-AI-L human annotation errors (mouse data).*

Comparison of initial manual annotations from iteration 0 (a and c) with their respective final network predictions from iteration 5 (b and d). These examples were selected due to poor manual annotation, where the glomerulus **(a)** was not annotated or **(c)** showed poorly drawn boundaries.

Network performance analysis is further complicated by human annotation errors. We note several instances where network predictions outperformed human annotators, despite being trained using flawed annotations. This phenomenon is highlighted in Figure 3, where glomerular regions annotated manually in iteration 0 are compared to the prediction by the iteration 5 network. Such errors are more prevalent in WSIs annotated in early iterations, where network predictions need the most correction.

To qualitatively prove the effectiveness and extendibility of our method, we show its extension to multi class detection by segmenting glomerular nuclei types [39, 40], interstitial fibrosis and tubular atrophy (IFTA) [41, 42], as well as differentiating sclerotic and non-sclerotic glomeruli [43] in mouse kidney and human renal biopsies. Figure 4 shows the glomeruli detection network from Figure 2 adapted for nuclei detection. This was done by re-training the high resolution network using a set of 143 glomeruli with labeled podocyte and non-podocyte nuclei, marked via immunofluorescence labeling. For this analysis, the low resolution network from Figure 2 was kept unchanged to identify the glomerular regions in the mouse WSI.

Due to the non-sparse nature of IFTA regions in some human WSI we forgo our DeepZoom approach to generate the results shown in Figure 5. The development of this IFTA network has been limited due to the biological expertise required to produce these multiclass annotations. However, preliminary segmentation results on holdout WSI show promising results despite using only 15 annotated biopsies for training (Figure 5). We note that this is a small training set, as human biopsy WSIs contain much less tissue area than the mouse kidney sections used to train the glomerular segmentation network above.

Finally, to show the adaptability of the H-AI-L pipeline to other medical imaging modalities, we quantify the use of our approach for the segmentation of human prostate glands from T2 MRI data. This data was orientated and normalized as described in [44] and saved as a series of TIFF image files, which can be opened in ImageScope and are compatible with our H-AI-L pipeline. This analysis was completed using a training set of data from 39 patients with an average of 32 slices per patient (512 x 512 pixels) (Figure 6d).

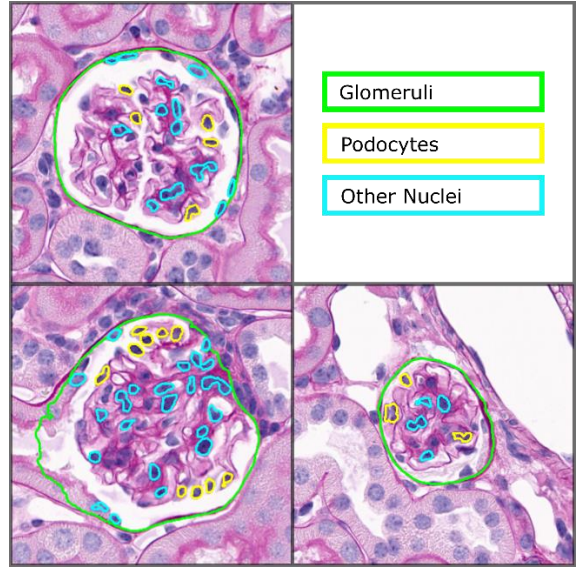

*Figure 4  Multiclass nuclei prediction on mouse WSI.*

Several examples of multiclass nuclei predictions are visualized on a mouse WSI. Here transfer learning was used to adapt the high resolution network from above (Figure 2) to segment nuclei classes. This network was trained using 143 labeled mouse glomeruli. The low resolution network was kept unchanged for the initial detection of glomeruli. We expect the results to significantly improve using more labeled training data.

Iterative training was completed by adding data from 4 patients to the training set prior to each iteration. Data from the remaining 7 patients was used as a holdout testing set. The newly annotated/corrected training data was augmented 10-times and a full resolution network was trained for 2 epochs during each iteration: the results of this training are presented in Figure 6. While the network performs well after just 1 round of training, the performance on holdout patient data continues to improve with the addition of training data (Figure 6a), achieving sensitivity of 0.88 ± 0.04, specificity of 0.99 ± 0.001, precision of 0.9 ± 0.03, and accuracy of 0.99 ± 0.001. This trend is also loosely reflected in the networks prediction on

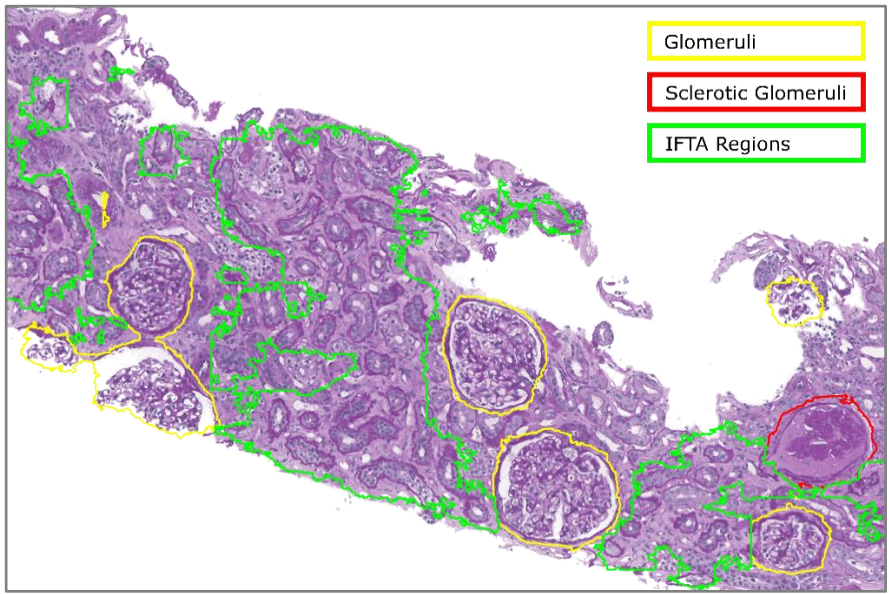

*Figure 5  Multiclass IFTA prediction on a holdout human renal WSI.*

Segmentation of healthy and sclerotic glomeruli, as well as IFTA regions from human renal biopsy WSI. Due to the non-sparse nature of IFTA regions, these predictions were made using only a high resolution pass. This is a screenshot of Aperio ImageScope which we use to interactively visualize the network predictions.

newly added training data, where an upward trend in prediction performance is observed in Figure 6b. Notably, when our iterative training pipeline is applied to this dataset, annotation is reduced by approximately 90 percent after the second iteration, where only 10 percent of MRI slices containing prostate fall below our segmentation performance threshold (Figure 6c).

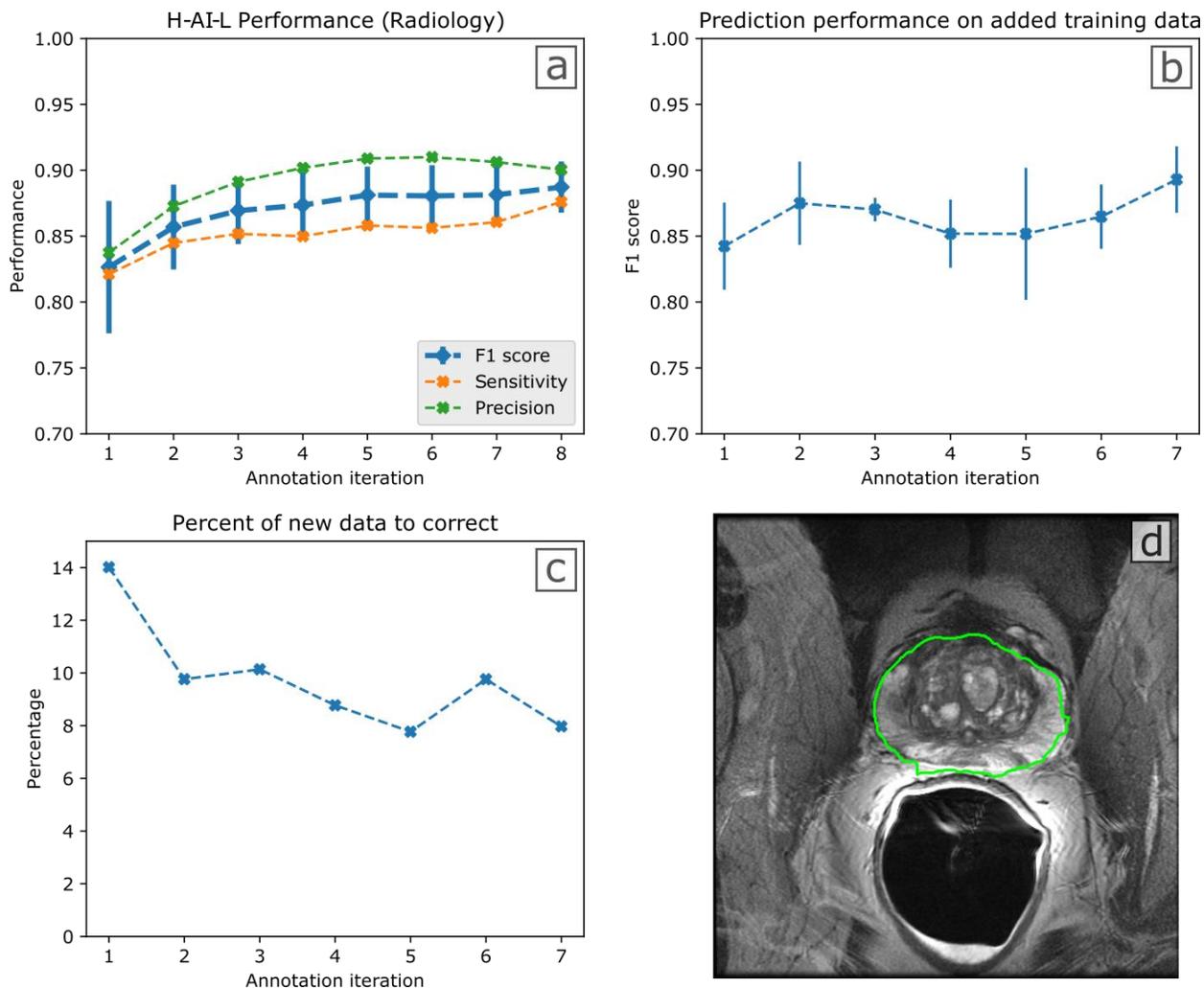

*Figure 6   H-AI-L method performance: human prostate segmentation from T2 MRI slices.*

**(a)** Segmentation performance as a function of training iteration, evaluated on 7 patient holdout MRI images (224 slices). Performance was evaluated on a patient basis. We note that despite the decline in network precision after iteration 6, the F1 score improves as a result of increasing sensitivity. **(b)** The prediction performance on added training data. This figure shows the prediction performance on newly added data w.r.t. the expert corrected annotation, and is evaluated on a patient basis (data from 4 new patients was added at the beginning of each training iteration). **(c)** The percentage of prostate regions where network prediction performance (F1 score) fell below an acceptable threshold (percentage of slices which needed expert correction) as a function of training iteration. We define acceptable performance as F1 score > 0.88. Using this criteria, expert annotation of new data is reduced by 92% by the fifth iteration. **(d)** A randomly selected example of a T2 MRI slice with segmented prostate: the network predictions are outlined in green. Error bars indicate ±1 standard deviation.

**Conclusions**

We have developed an intuitive pipeline for segmentation of structures from WSI commonly used in pathology, a field where there is often a large disconnect between domain experts and engineers. We aim to bridge this gap by making the robust data analytics provided by state-of-the-art neural networks accessible to pathologists. Along this direction we have developed an intuitive library for the adaptation of DeepLab v2 [26], a semantic segmentation network, to whole slide imaging data, commonly used in the field. This library uses annotation tools from the common WSI viewing software Aperio ImageScope [27] for annotation and display of the network predictions. Training, prediction and validation of the network is done via a single python script with a command line interface, where data management is as simple as dropping data into a pre-determined folder structure.

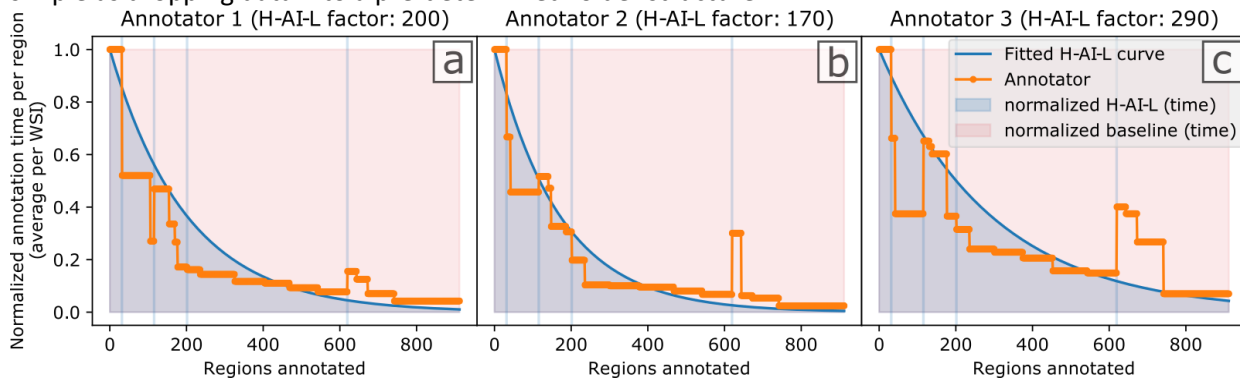

*Figure 7  Annotation time savings using the H-AI-L method: compared to baseline segmentation speed (Figure 2a).*

H-AI-L plots showing the annotation time per region normalized with respect to the baseline annotation speed of each annotator. An exponential decay distribution (H-AI-L curve) is fitted to each annotator, where the H-AI-L factor is the exponential time constant: a derivation can be found in the methods section. The vertical lines are gaps between iterations (where the network was trained). The area under the H-AI-L curve represents the normalized annotation time per annotator. This can be compared to the area of the normalized baseline region, which represents the normalized annotation time without the H-AI-L method. **(a)** The time savings by annotator 1 (calculated to be 81.3 percent) when creating the training set used to train the glomerular segmentation network in Figure 2. **(b)** Annotator 2 was 82.0 percent faster. **(c)** Annotator 3 was 72.7 percent faster. While the y-axis in these plots is not a direct measure of network performance, it is highly correlated. The spike in annotation time seen at 600 regions is data from a WSI with severe glomerular damage from DN. We believe that plots like these will offer insight into optimal iterative training strategies in the future, with a goal of reducing annotation burdens for expert annotators.

Using our iterative, human in the loop training allows considerably faster annotation of new WSIs (or similar imaging data), as network predictions can easily be corrected in ImageScope before incorporation into the training set. This approach allows the qualitative assessment of network performance after each iteration, as newly added data acts as a holdout validation set, where predictions are easily viewed during correction. The theoretical performance achievable by this method is bounded by the training set used, and is therefore the same as the current state-of-the-art (manual annotation of all training data). However, due to the increased speed of annotation, and intuitive visualization of network performance (allowing selection of poorly predicted new data after each iteration), we argue that H-AI-L training has the potential to converge to the upper bound of performance more efficiently than the traditional method; achieving state-of-the-art segmentation performance much faster than traditional methods, which are limited by data annotation speed (Figure 7). To our knowledge, our approach: displaying network predictions in ImageScope, is the first of its kind. It offers an ideal viewing environment for network predictions on WSIs, using the fast pan and zoom functionality provided by ImageScope [28], improving the accuracy and ease of expert annotation.

The ability to transfer parameters from a trained network (repurposing it for a different task), ensures that segmentation of tissue structure can be tailored to any clinical or research definition, including other biomedical imaging modalities. Our multiresolution (DeepZoom) analysis allows rapid prediction of sparse regions from large WSIs, without sacrificing accuracy due to low resolution analysis alone. Inspired by the way pathologists scan tissue slides, multiresolution approaches have been successfully used in digital pathology literature for the detection of cell nuclei [45]. We believe that this technique offers the perfect compromise between speed and specificity, producing high resolution sparse segmentations ideal for display in ImageScope. The use of our method for non-sparse segmentation of WSI is achievable by foregoing DeepZoom analysis. However, in the future we plan to change the way that the class hierarchy is defined in our algorithm, offering easy functionality to search for low resolution regions with high resolution sub-compartments.

In the future we will undergo extensive testing of our method in a clinical research setting. This testing would involve evaluation of the segmentation performance as well as ergonomic aspects which pertain to a clinician's ease of use. We will extend our method to provide anomaly detection, defining a confidence metric and threshold where WSIs are flagged for further evaluation. To compliment this, we will create an algorithm to predict the optimal amount of annotation in each iteration (to optimize expert time) using a curve fitting similar to figure 6. We will also adapt our method for native use with a DICOM viewer, allowing easier workflows for segmentation of Radiology datasets. Given these tools, we foresee a segmentation approach similar to our H-AI-L method acting as a cornerstone of efforts to build searchable databases of digital pathology slides [22], and other medical imaging datasets.

**Methods**

All animal tissue sections were collected in accordance with protocols approved by the Institutional Animal Care and Use Committee at University at Buffalo, and are consistent with federal guidelines and regulations and in accordance with recommendations of the American Veterinary Medical Association guidelines on euthanasia. Renal biopsy samples were collected from the Kidney Translational Research Center at Washington University School of Medicine, directed by co-author Dr. Jain, following a protocol approved by the Institutional Review Board at University at Buffalo prior to commencement. Digital MRI images of human prostate glands were provided by co-author Dr. LaViolette, following a protocol approved by the Institutional Review Board at Medical College of Wisconsin. All methods were performed in accordance with the relevant federal guidelines and regulations. All patients provided written informed consent, and basic demographic information was collected.

In the H-AI-L pipeline, an annotator labels one whole slide image using annotation tools in ImageScope [27], which provides the input for network training. The resulting trained network is then used to predict the annotations on a new WSIs. These predictions are used as rough annotations, which are corrected by the annotator and sent back for incorporation into the training set; improving network performance and optimizing the amount of expert annotation time required. Because this technique makes the adaptation of network parameters to new data easy, adapting a trained network to new data generated in different institutions is extremely feasible. We have made our code openly available online: https://github.com/SarderLab/H-AI-L

At the heart of H-AI-L is the conversion between mask and XML [46] formats which are used by DeepLab v2 [26] and ImageScope [27], respectively. Training any semantic segmentation architecture relies on pixel-wise image annotations which are input to the network for training and output after network predictions as mask images. In the case of DeepLab, the mask images take the form of indexed greyscale 8 bit PNG files, where each unique value pertains to an image class. On the other hand, annotations done in ImageScope are saved in text format, as XML files [46], where each region is saved as a series of boundary points or vertices. Determining the vertices of a mask image is a common image processing task, known as image contour detection [47, 48]. As opposed to edge detection, contour detection can have hierarchal classifications [48], lending itself ideally to conversion into the hierarchal XML format used by ImageScope.

To facilitate the transfer between ImageScope XML and greyscale mask images, we use the *OpenCV-Python* library (*cv2*) [47], using the function *cv2.findContours* to convert from masks to contours. Using this function, we are able to automatically convert DeepLab predictions to XML format which can be viewed in ImageScope, easily evaluating and correcting network performance. Additionally, we have written a library for converting an XML file into mask regions, using *cv2.fillPoly.* This library follows the *OpenSlide-Python* [49] conventions for reading WSI regions, returning a specified mask region from the WSI.

OpenSlide [49] and our XML to mask libraries allow for efficient chopping of WSI into overlapping blocks for network training and prediction; similar sliding window approaches are common practice for predicting semantic segmentations on large medical images [50, 51]. To simplify the iterative training process, and compliment the easy annotation pipeline proposed, we have created a callable function which handles operations automatically, prompting the user to initiate the next step. This function needs two flags [--*option*] and [--*project*] which are the parameters identifying the iterative step and project one would like to train respectively. Initially created using [--*option*] *'new'*, a new project is trained iteratively by alternating the [--*option*] flag between *'train'* and *'test'*. Our algorithm uses our DeepZoom approach by default, but full-resolution analysis is achievable by setting the [--*one_network*] flag to *'True'* during training and prediction.

*Training:*

To streamline the training process, we created a pipeline where a user places new WSIs and XML annotations in a project folder structure, then calls a function to train the project. This automatically initiates data chopping and augmentation, then loads parameters from the most recently trained network (if available) before starting to train. For faster convergence, we utilize transfer learning, automatically pulling a pre-trained network file whenever a new project is created, which is used to initialize the network parameters prior to training. We have also included functionality to specify a pre-trained file from an existing project using the [--*transfer*] flag. For ease of use, the network hyper-parameters can be changed using command line flags, but are set automatically by default.

When [--*option*] *'train'* is specified, WSIs and XML annotations are chopped into a training set containing 500 x 500 blocks with 50% overlap. This training set is then augmented via: random flipping, hue and lightness shifts, as well as piecewise affine transformations; accomplished using the *imgaug* python library [52]. To keep the network unbiased, the total number of blocks containing each class is tabulated and used to augment less frequent classes with a higher probability [53]. Once augmented, the network is trained for the specified number of epochs, and the user is prompted to upload new WSIs and run the [--*option*] *'predict'* flag. This produces XML predictions which can be corrected using ImageScope before incorporation into the training set.

*Prediction:*

Due to the sparse nature of the structures we attempt to segment from renal WSI, we limit the search space, using a low resolution pass to determine hotspot regions before segmentation at full resolution (DeepZoom). This is accomplished in two ways: Firstly, thresholding and morphological processing are used to determine which WSI blocks contain tissue, eliminating background regions. Secondly, down-sampled blocks (1/16th resolution) are tested using a semantic segmentation network (DeepLab) to roughly segment structures. The output predictions of the preprocessing steps are then stitched back into a hotspot map, which identifies important regions at this resolution. Using this map, full size hotspot indices are calculated, and the regions are extracted using OpenSlide for pixel-wise segmentation by a second network.

*Validation:*

While the performance of network is easily visualized after prediction on new WSI, we have included functionality for explicit evaluation of performance metrics and prediction time on a holdout dataset. This is accomplished using the [--*option*] *'validate'* flag. When called, it evaluates the network performance on holdout images for every annotation iteration by pulling the latest models automatically. To perform this performance comparison, ground truth XML annotations of the holdout set are required for the calculation of sensitivity, specificity, accuracy, and precision performance metrics [37].

*Estimating H-AI-L performance (Figure 7):*

To quantify the time savings of our H-AI-L method, we plot the normalized annotation time per region vs the number of regions annotated. Here we define the normalized annotation time per region $A$ as:

$$A = \frac{t}{t_0},$$

where $t$ is the annotation time per region (averaged per WSI), and $t_0$ is the average annotation time per region in iteration 0. $A$ is bounded from $[0,1]$ where 1 is the normalized time it takes to annotate one region fully. While the annotation time is reduced as a piecewise function of training iteration, in Figure 7 we use a continuous exponential decay distribution to approximate $A(r)$:

$$A(r) = e^{-\frac{r}{\tau}},$$

where $r$ is the number of regions annotated, and $\tau$ is the exponential time constant which we call the H-AI-L factor.

The normalized annotation time of our H-AI-L method ($H$) can therefore be estimated as:

$$H = \int_0^R A(r)\mathrm{d}r = \tau\left[1 - e^{\frac{-R}{\tau}}\right],$$

where $R$ is the total number of regions annotated. Like-wise, the normalized baseline annotation time ($B$) can be calculated as:

$$B = \int_0^R 1 \mathrm{d}r = R$$

Therefore the time savings performance ($P$) of our H-AI-L method can be estimated as a percentage using:

$$P = \left(1 - \frac{H}{B}\right) * 100 = \left(1 + \frac{\tau}{R}\left[e^{\frac{-R}{\tau}} - 1\right]\right) * 100.$$

The H-AI-L factor $\tau$ reflects the effectiveness of iterative network training, where lower values of $\tau$ represent training curves that decay faster. In the future, algorithms to select the optimal amount of annotation and identify data outliers to be annotated at each iteration will improve the performance of the H-AI-L method by reducing $\tau$.

**Acknowledgements**

The project was supported by the faculty startup fund from the Pathology and Anatomical Sciences Department, University at Buffalo, the University at Buffalo IMPACT grant, the DiaComp Pilot and Feasibility Program grant #32307-5, and NIDDK grant R01 DK114485. We thank NVIDIA Corporation for the donation of the Titan X Pascal GPU used for this research (NVIDIA, Santa Clara, CA).

**Author contributions**

B.L. conceived the H-AI-L method, analyzed the data, and wrote the paper. The code was written by B.L. and B.G. D.G. contributed in generating results for Figure 4. S.D.M. and P.S.L. provided the radiology data and annotations for Figure 6. R.Y. implemented the mouse model. S.J. provided human renal biopsy data. J.E.T. evaluated renal pathology segmentation as a domain expert. K.Y.J. provided the IFTA annotation for Figure 5. P.S. is responsible for the overall coordination of the project, mentoring and formalizing the image analysis concept, and oversaw manuscript preparation.

**Competing interests**

The authors declare they have no competing interests.

*Supplemental Figure 1      Network prediction evolution*

H-AI-L network predictions visualized in 1/10[th] Epoch steps: the training iteration is visualized in the top right corner of the images for reference. New training data is added each training iteration: we note the large jumps in performance achieved quickly at the start of each iteration.

**This figure is a video file, available from: https://buffalo.box.com/s/7lahd5tz1zyc3xy7ad1axmfrdfei09ln**

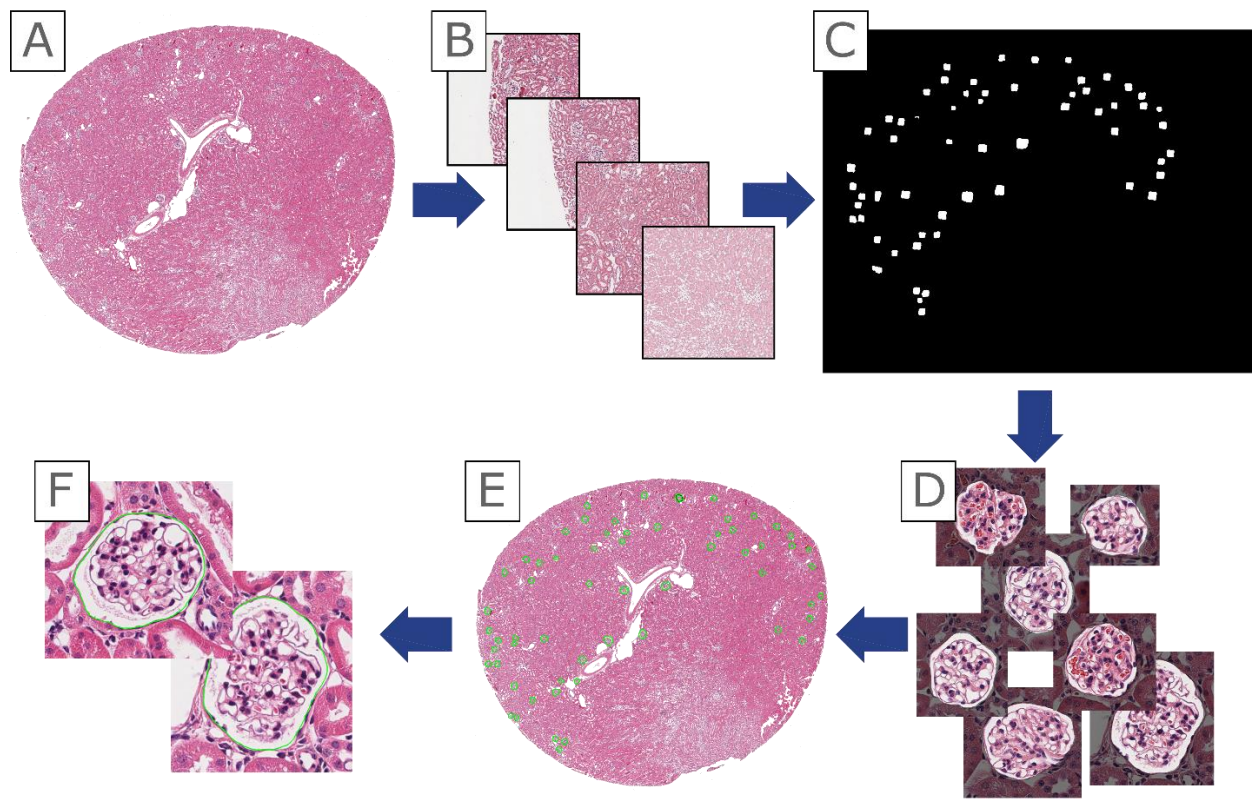

*Supplemental Figure 2      Outline of the DeepZoom H-AI-L method for semantic segmentation of WSIs*

An overview of our H-AI-L and DeepZoom approach for fast sparse semantic segmentation of WSI using convolutional neural networks. **(A)** shows the WSI, **(B)** the extraction of low resolution ($1/16^{th}$x) blocks by a sliding window, **(C)** the initial low resolution predictions on the blocks stitched into a hotspot map, **(D)** the extracted hotspot regions, segmented by the high resolution network, **(E)** the final predictions displayed on the WSI in ImageScope via conversion to XML, **(F)** a closer view of the predictions displayed in ImageScope, demonstrating the zoom functionality of ImageScope. This using this pipeline, fast sparse semantic segmentation of WSI is achievable, with easily viewable predictions.